\documentclass[aps,prl,twocolumn,showpacs,superscriptaddress]{revtex4}
\usepackage{graphics,bm}
\usepackage{amssymb}
\usepackage{epsfig}
\usepackage{epsf}

\def\Re{{\rm Re}}
\def\Im{{\rm Im}}
\def\be{\begin{equation}} \def\ee{\end{equation}}
\def\beq{\begin{eqnarray}} \def\eeq{\end{eqnarray}}

\def\nn{\nonumber}

\begin{document}

\title{Resonant Plasmon-Axion Excitations Induced by Charge Density Wave Order in Weyl Semimetal}

\author{Matthew D. Redell}
\affiliation{Department of Physics, Applied Physics, and Astronomy, Binghamton University - State University of New York, Binghamton, USA}

\author{Shantanu Mukherjee}
\affiliation{Department of Physics, Applied Physics, and Astronomy, Binghamton University - State University of New York, Binghamton, USA}

\author{Wei-Cheng Lee}
\email{wlee@binghamton.edu}
\affiliation{Department of Physics, Applied Physics, and Astronomy, Binghamton University - State University of New York, Binghamton, USA}

\date{\today}

\begin{abstract}
We investigate the charge excitations of a Weyl semimetal in the axionic charge density wave (axionic CDW) state. 
While it has been shown that the topological response (anomalous Hall conductivity) is protected against the CDW state, we find that the long wavelength plasmon excitation is 
radically influenced by the dynamics of the CDW order parameter.
In the normal state, we show that an undamped collective mode should exist at $\vec{q}\approx \vec{Q}_{CDW}$ if there is an attractive interaction favoring the formation of the CDW state.
The undamped nature of this collective mode is attributed to a gap-like feature in the particle-hole continuum at $\vec{q}\approx \vec{Q}_{CDW}$ due to the chirality of the Weyl nodes, which
is not seen in other materials with CDW instability.
In the CDW state, the long wavelength plasmon excitations become more dispersive due to the additional interband scattering not allowed in the normal state.
Moreover, because the translational symmetry is spontaneously broken, 
Umklapp scattering, the process conserving the total momentum only up to $n\vec{Q}_{CDW}$ with $n$ an integer and $\vec{Q}_{CDW}$ the ordering wave vector, emerges in the CDW state. 
We find that the plasmon excitation couples to the phonon mode of the CDW order via the Umklapp scattering, leading to two branches of resonant collective modes 
observable in the density-density correlation function at $\vec{q}\approx 0$ and $\vec{q}\approx \vec{Q}_{CDW}$.
Based on our analysis, we propose that measuring these resonant plasmon-axion excitations around $\vec{q}\approx 0$ and $\vec{q}\approx \vec{Q}_{CDW}$ by the momentum-resolved 
electron energy loss spectroscopy (M-EELS) could serve as a 
reliable way to detect the axionic CDW state in Weyl semimetals.

\end{abstract}


\maketitle

{\it Introduction}--
The Weyl semimetal\cite{wanx2011,burkov2011} is a new gapless state of matter attracting a lot of attention in the past few years due to its rich 
topological properties.\cite{yangk2011,vivek2012,zyuzin2012,hosur2012,goswami2013,vazifeh2013,liuc2013,son2013,panfilov2014,burkov20141,burkov20142,bulmash2015}
In the wake of the study on topological insulators\cite{hasan2010,qi2011}, the search for semimetals exhibiting linear band dispersion near the Fermi energy has been one of the most active fields 
in condensed matter physics. One typical example is the Dirac semimetal\cite{wangzh2012,wangzh2013,liuz2014,borisenko2014} in which the conduction and valence bands 
become degenerate at certain points in the Brillouin zone, namely the Dirac nodes, and 
the low energy single particle states near the Dirac nodes can therefore be described by Dirac equations with a four-component spinor. 
If the Dirac semimetal has a zero mass, the Dirac equation can be further reduced to two independent sets of Weyl equations with two-component spinors. 
As a result, the Weyl semimetal is a special case of the Dirac semimetal with zero mass, which could be obtained by breaking either the inversion or the time-reversal symmetries in a Dirac semimetal.
In addition, it can be proved\cite{nielsen19811,nielsen19812,nielsen1983} that Weyl nodes must appear in pairs with opposite chirality if the Weyl equations are imposed on a lattice.
Therefore, if a crystalline material is a Weyl semimetal, the low energy single particle excitations can be described by a minimal model of the Weyl fermions 
around two Weyl nodes obeying the Weyl equations:
\be
H_0= \hbar v_F\int d\vec{k} \sum_{\chi=\pm 1} \chi\vec{k}\cdot\vec{\sigma} - \mu_{\chi}\hat{I},
\label{h0}
\ee
where $v_F$ is the Fermi velocity, and $\chi=\pm 1$ is the chirality of the Weyl nodes.
Very recently, the experimental realization of the Weyl semimetal has been found in TaAs.\cite{xus2015,wengh2015}

The spontaneous symmetry breaking phases in the Weyl semimetal induced by the electron-electron correlation have also been investigated.\cite{weih2012,wangz2013,maciejko2014,weih2014,sekine2014,bednik2015,bitan2015,zhangr2016}
One intriguing aspect unique to the Weyl semimetal is the chiral anomaly accompanying the charge-density-wave (CDW) state.\cite{wangz2013,bitan2015}
If the pair of the Weyl nodes described in Eq. \ref{h0} is located at $\pm \vec{Q}$ in the Brillouin zone, the electron-electron interaction could result in a CDW state 
with an ordering wavevector $\vec{Q}_{CDW} = 2\vec{Q}$. Since in the CDW state the electron occupation number at $\pm\vec{Q}$ is different, the chiral symmetry is also broken.
It can be further shown that the phase of the CDW order parameter can be identified as the 'axion field' which couples to the electromagnetic fields and exists only in the presence of the chiral 
anomaly.\cite{wangz2013,bitan2015}

In this paper, we study the charge excitations of the Weyl semimetal in the axionic CDW state. In the normal state, we find an undamped collective mode at $\vec{q}\approx \vec{Q}_{CDW}$ 
if there is an attractive interaction favoring the formation of the CDW state. In the CDW state,
we find that the nature of the long wavelength plasmon excitation is dramatically changed. The phonon mode of the CDW order, or equivalently the collective mode of the axion field, couples to 
the plasmon excitation via the Umklapp process, resulting in two resonant collective modes which can be seen in the density-density correlation function at both $\vec{q}\approx 0$ and 
$\vec{q}\approx \vec{Q}_{CDW}$. 
This resonant feature does not emerge in the case where the chiral symmetry is broken by the application of a pairing of the electric $\vec{E}$ and the magnetic $\vec{B}$ fields.\cite{hosur2015,zhouj2015}
We propose that measuring these resonant plasmon-axion excitations around $\vec{q}\approx 0$ and $\vec{q}\approx \vec{Q}_{CDW}$ by the momentum-resolved electron energy loss spectroscopy
(M-EELS)\cite{kogar2014} could serve as a reliable way to detect the CDW state as well as
the nature of the axion field in a Weyl semimetal.

{\it Hamiltonian and Formalism} --
We start from the effective Hamiltonian $H_0$ describing the Weyl fermions around two Weyl nodes at $\pm\vec{Q}$ given in Eq.\ref{h0} together with electron-electron interactions:
\beq
H&=&H_0 + H_{CDW} + H_{LC},\nn\\
H_{CDW} &=& V_{CDW}\int d\vec{q}  \rho(\vec{q}+\vec{Q}_{CDW})\rho(-\vec{q}-\vec{Q}_{CDW}),\nn\\
H_{LC} &=& \int d\vec{q} v_{\vec{q}}\rho(\vec{q})\rho(-\vec{q}).
\eeq
where $v_{\vec{q}} = 4\pi e^2/\kappa q^2$ is the 3D Coulomb interaction in momentum space. 
It has been shown in previous study that the CDW instability is one of the leading instabilities in the Weyl semimetal induced by repulsive interactions\cite{weih2012,maciejko2014,bitan2015},
which justifies using $H_{CDW}$ to stduy the CDW fixed point.
The charge density operator can be expanded around the proximity of the Weyl nodes, leading to
\beq
&&\rho(\vec{q}) \approx \sum_\sigma \int^\Lambda d\vec{k} c^\dagger_{R\sigma}(\vec{k} + \vec{q}) c_{R\sigma}(\vec{k})+c^\dagger_{L\sigma}(\vec{k}) c_{L\sigma}(\vec{k} - \vec{q}),\nn\\
&&\rho(-\vec{q}-\vec{Q}_{CDW}) \approx \sum_\sigma \int^\Lambda d\vec{k} c^\dagger_{L\sigma}(\vec{k} - \vec{q}) c_{R\sigma}(\vec{k})
\label{den}
\eeq
where $\vec{k}$ are small momenta compared to $\vec{Q}$, $\Lambda$ is the cut-off in the integration over the momentum, and $\rho(\vec{q}+\vec{Q}_{CDW}) = \rho^\dagger(-\vec{q}-\vec{Q}_{CDW})$.
Since we are only interested in the long-wavelength plasmon excitations and the charge fluctuations near the CDW ordering wave vector $\vec{Q}_{CDW}$,
the integration over $\vec{q}$ is limited to the region of small $\vert\vec{q}\vert$

We perform the standard mean-field theory to decouple $H_{CDW}$ with the axionic CDW order parameter defined as $m_1 - i m_2\equiv \langle \rho(\vec{Q}_{CDW})\rangle$, 
and consequently the mean-field Hamiltonian in the CDW state can be written as\cite{wangz2013,bitan2015}
$H^{CDW}_{MF} = \hbar v_F\int d\vec{k} \hat{\psi}^\dagger(\vec{k}) \hat{h}(\vec{k}) \hat{\psi}(\vec{k})$,
where 
\beq
\hat{h}(\vec{k}) &=& 
\left(
\begin{array}{cc}
\vec{k}\cdot\vec{\sigma} - \mu_+\hat{I}&\hat{m}^\dagger\\
\hat{m}&-\vec{k}\cdot\vec{\sigma} - \mu_-\hat{I}
\end{array}
\right),\nn\\
\hat{m} &=&
\left(
\begin{array}{cc}
m_1 - im_2&0\\
0&m_1 -im_2
\end{array}
\right),
\eeq
, and $\hat{\psi}^\dagger(\vec{k}) = \big[c^\dagger_{R\uparrow}(\vec{k}), c^\dagger_{R\downarrow}(\vec{k}), 
c^\dagger_{L\uparrow}(\vec{k}), c^\dagger_{L\downarrow}(\vec{k})\big]$. $c^\dagger_{R\sigma}(\vec{k})$ and $c^\dagger_{L\sigma}(\vec{k})$ are the creation operators near $+\vec{Q}$ and $-\vec{Q}$ 
respectively, and they are related via $c^\dagger_{L\sigma}(\vec{k}+\vec{Q}_{CDW}) = c^\dagger_{R\sigma}(\vec{k})$, where $\vec{Q}_{CDW} = 2\vec{Q}$ is the CDW ordering wave vector. 
$H^{CDW}_{MF}$ can be diagonalized as $H^{CDW}_{MF} = \hbar v_F\int d\vec{k} \hat{\alpha}^\dagger(\vec{k}) \hat{h}^D(\vec{k}) \hat{\alpha}(\vec{k})$,
where $\hat{h}^D={\rm Diag}(E_1(\vec{k}),E_2(\vec{k}),E_3(\vec{k}),E_4(\vec{k}))$, 
and $\hat{\alpha}^\dagger=\big[\alpha_1(\vec{k}),\alpha_2(\vec{k}),\alpha_3(\vec{k}),\alpha_4(\vec{k})\big]$ are the 
corresponding eigenvectors which are related to $\psi(\vec{k})$ via a unitary matrix $\hat{U}$ such that $\hat{\psi}(\vec{k}) = \hat{U}(\vec{k}) \hat{\alpha}$.

The CDW order parameter is self-consistently obtained by solving the mean-field equation:
\be
\frac{m_1 - i m_2}{ -V_{CDW}} = \int d\vec{k} \sum_\sigma U^*_{R\sigma,\alpha}(\vec{k})U_{L\sigma,\alpha} n_F(E_\alpha),
\ee
which supports a non-zero order parameter if $V_{CDW}$ is negative and smaller than a critical value ($V_{CDW} <  V^c_{CDW} < 0$).

Clearly, in the normal state, the density fluctuations at $\vec{q}+\vec{Q}_{CDW}$ are independent of those at $\vec{q}$, 
thus they do not affect the long wavelength plasmon excitations at all. If the system is in the CDW state, a non-zero order parameter
induces a coupling between the density fluctuations at $\vec{q}$ and $\vec{q}+\vec{Q}_{CDW}$. Such a coupling can be described by the Umklapp scattering in which the total momentum is
conserved only up to $n\vec{Q}_{CDW}$ with $n$ an integer, and significant changes in the properties of the long wavelength plasmon excitations occurs. 
Note that in principle all the density fluctuations in different Umklapp channels (i.e., different $n$) are coupled, but for the demonstration of principle, it is sufficient to consider the 
two channels given in Eq. \ref{den}.

To take the Umklapp process into account, we introduce a matrix form to represent the density-density correlation function written as
$\hat{\chi}_{ij}(\vec{q},\omega)\equiv \langle\rho_i(\vec{q},\omega) \rho^\dagger_j(\vec{q},\omega)\rangle$,
where $\rho_1(\vec{q},\omega)$ and $\rho_2(\vec{q},\omega)$ refer to $\rho(\vec{q},\omega)$ and $\rho(\vec{q}+\vec{Q}_{CDW},\omega)$ respectively.
In general, $\hat{\chi}_{11}$ refers to the charge excitations inside each Weyl nodes, and $\hat{\chi}_{22}$ refers to those between different Weyl nodes.
Performing a generalized RPA which has been widely used in studying the coupling between different collective excitations\cite{grpa1,grpa2,grpa3,grpa4,grpa5}, we obtain
\be
\hat{\chi}^{GRPA}(\vec{q},\omega) = \big[ \big(\hat{\chi}^0(\vec{q},\omega)\big)^{-1} + \hat{V}(\vec{q})\big]^{-1}
\ee
where $\hat{\chi}^{GRPA}$, $\hat{\chi}^0$, and $\hat{V}(\vec{q})$ are $2\times 2$ matrices,
\beq
\hat{\chi}^0_{i,j}(\vec{q},\omega) &=& - \sum_{\alpha,\beta=1}^4 \int^\Lambda d\vec{k} M_{i,j}^{\alpha,\beta}(\vec{k},\vec{q})\nn\\
&\times&\big(\frac{n_F(E_\alpha(\vec{k}+\vec{q}))-n_F(E_\beta(\vec{k}))}{\hbar\omega + i\delta
+E_\alpha(\vec{k}+\vec{q}) - E_\beta(k)}\big),\nn\\
\label{chi0}
\eeq
$M_{i,j}^{\alpha,\beta}(\vec{k},\vec{q})$ is the matrix element which is given in the Supplementary Materials, and
$\hat{V}_{\vec{q}}={\rm Diag.}(v_{\vec{q}},V_{CDW})$ is the interaction kernel.
%\be
%\hat{V}_{\vec{q}}=
%\left(
%\begin{array}{cc}
%v_q&0\\
%0&V_{CDW}
%\end{array}\right)
%\ee
For the calculations, we adopt the Lorentz cut-off, i.e., $\vert \vec{k}\vert < \Lambda$ and $\omega < v_F \Lambda$. Since we only focus on the excitations near the Weyl nodes, 
$q << \Lambda < \vert\vec{Q}_{CDW}\vert$. 
Furthermore, we choose $k_F$ and $E_0 = \hbar v_F k_F$ to be the units of momentum and energy in all of our calculations.

\begin{figure}
\includegraphics[width=2.9in]{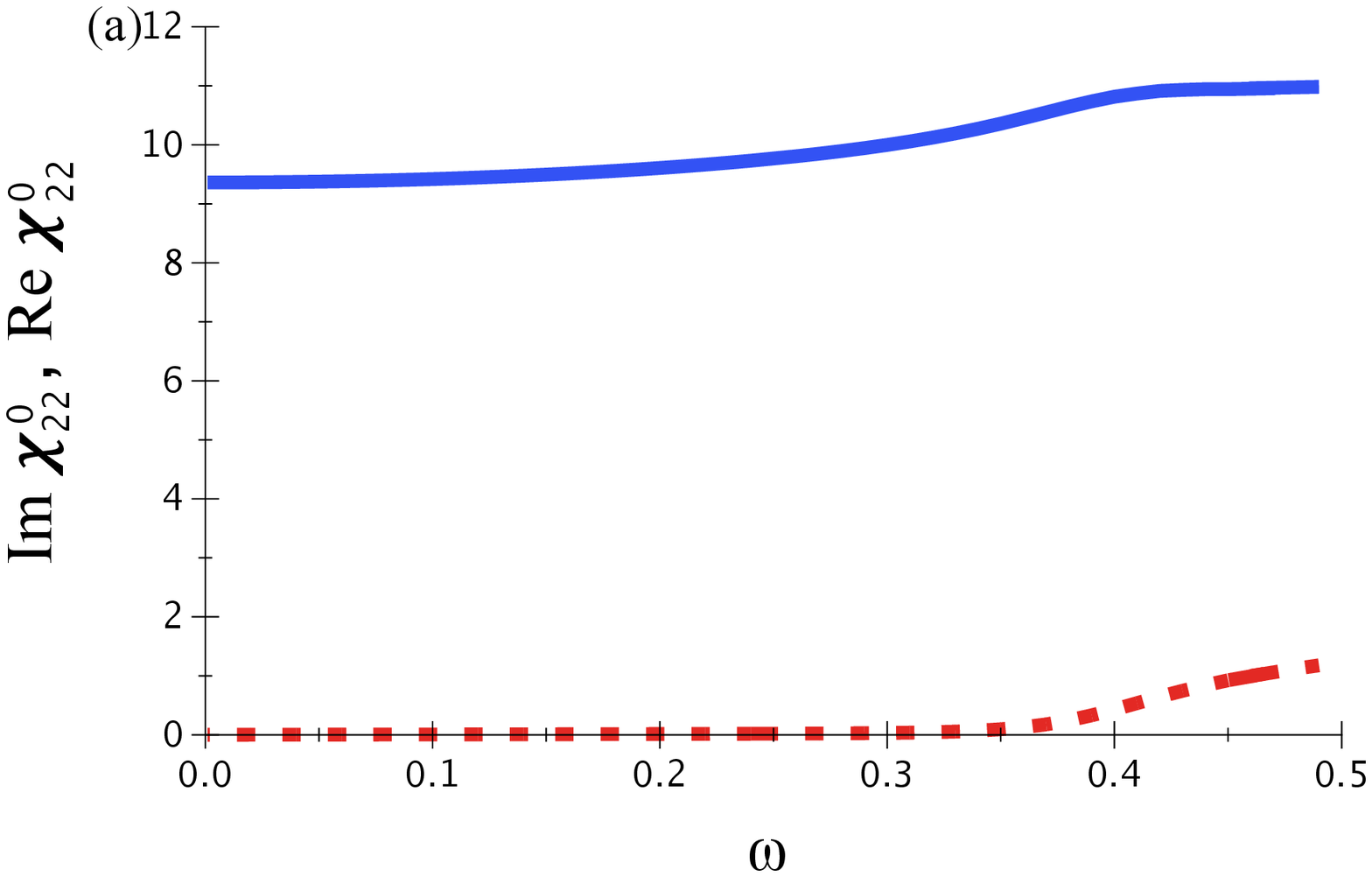}
\includegraphics[width=2.9in]{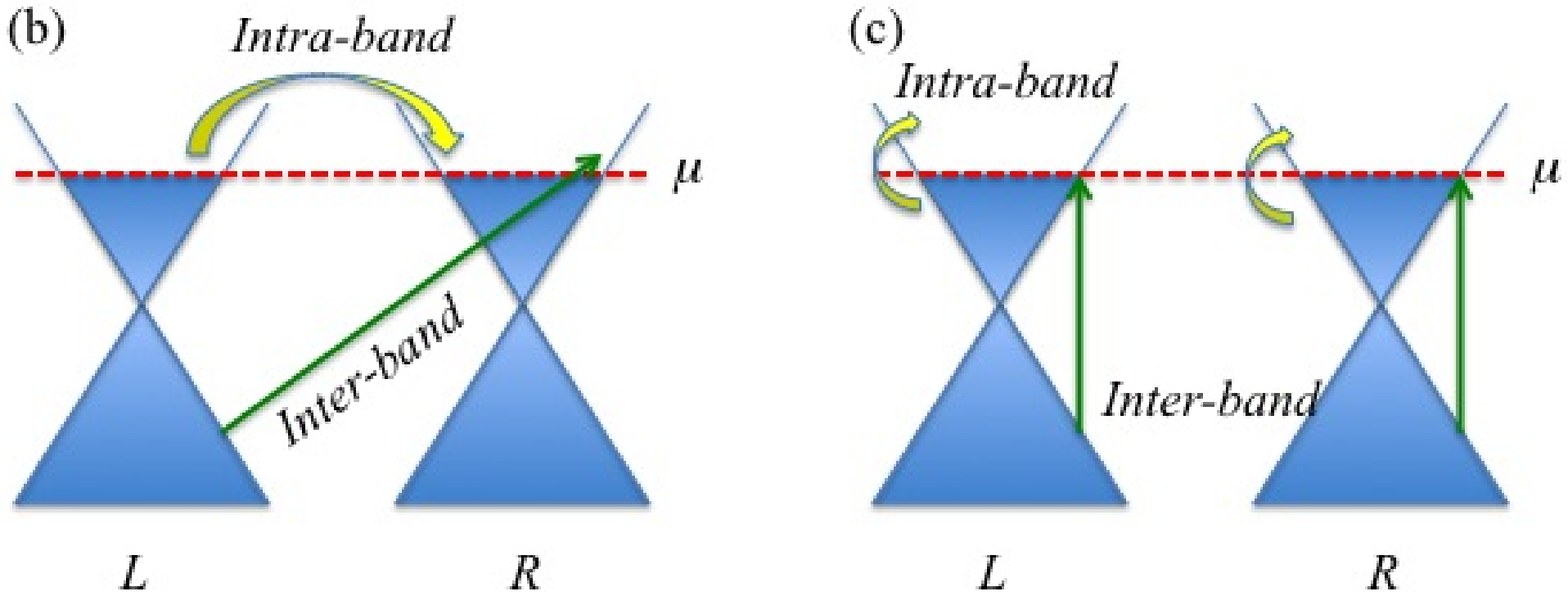}
\caption{\label{fig:chi022} (a) Real (blue solid) and imaginary (red dashed) parts of $\chi^0_{22}(\vert\vec{q}\vert=0.01,\omega)$. The parameters used are $\mu_{\pm} = 0.2$, $m_1 = m_2 = 0$
and $\Lambda=5\mu$.
The particle-hole excitations from the intraband scatterings are greatly suppressed, resulting
a gap-like feature in $\Im\chi^0_{22}(\vert\vec{q}\vert=0.01,\omega)$. The intraband and interband scatterings in $\chi^0_{22}(\vec{q},\omega)$ and
$\chi^0_{11}(\vec{q},\omega)$ are schematically illustrated in (b) and (c) respectively.}
\end{figure}

\begin{figure}
\includegraphics[width=2.9in]{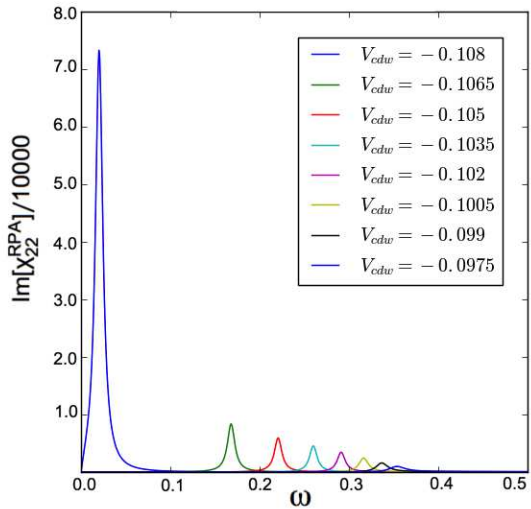}
\caption{\label{fig:chirpa22} The undamped collective mode in $\Im\chi^{RPA}_{22}(\vec{q}\approx 0,\omega)$ associated with the CDW instability in the normal state. The peak moves toward zero frequency 
as $V_{CDW}$ approaches $V^c_{CDW}$. The parameters used are $\mu_{\pm} = 0.2$, $m_1 = m_2=0$ and $\Lambda=5\mu$.}
\end{figure}

{\it Collective charge excitation in normal state} --
In the normal state without CDW, $\chi^0_{12}(\vec{q},\omega) = 0$. As a result, $\chi^{RPA}_{11}(\vec{q},\omega)$
and $\chi^{RPA}_{22}(\vec{q},\omega)$ are decoupled. 
$\chi^{RPA}_{11}(\vec{q},\omega)$ has an undamped delta-function-like peak corresponding to the plasmon excitations, 
which has been studied previously\cite{lvm2013,panfilov2014,hofmann2015}. 
As for $\chi^{RPA}_{22}(\vec{q},\omega)$, we find that an undamped delta-function-like peak could emerge if $V_{CDW}$ is negative. 
The occurrence of this undamped mode is due to a gap-like feature appearing in the bare susceptibility $\chi^0_{22}(\vec{q},\omega)$ shown in Fig. \ref{fig:chi022}(a). 
Because the Weyl nodes have opposite chirality, the intraband scattering is largely suppressed while the interband scattering is robust in $\chi^0_{22}(\vec{q},\omega)$ as demonstrated 
in Fig. \ref{fig:chi022}(b). This unusual chirality effect leads to $\Im\chi^0_{22}(\vec{q},\omega) \approx 0$ for $\omega < \omega_{PHE_{inter}}$, where $\omega_{PHE_{inter}} = 2\mu - v_F q$ 
is the particle-hole continuum edge for the interband scattering. 
In contrast, $\chi^0_{11}(\vec{q},\omega)$ is dominated by the intraband scattering as sketched in Fig. \ref{fig:chi022}(c), 
and consequently $\Im\chi^0_{11}(\vec{q},\omega)$ is finite for $\omega < \omega_{PHE_{intra}}$, where $\omega_{PHE_{intra}} = v_F q$ is the particle-hole continuum edge for the intraband scattering, 
and almost zero for $\omega > \omega_{PHE_{inter}}$. 
By Kramers Kronig relations, it can be proved that $\Re\chi^0_{22}(\vec{q},\omega)$ increases monotonically from $\omega=0$ until $\omega = \omega_{PHE_{inter}}$. 
If $V_{CDW}$ is negative, a collective mode with the frequency $\Omega$ satisfying $1+V_{CDW}\chi^0_{22}(\vec{q},\Omega)=0$ exists, which is corresponding to 
the undamped delta-function-like peak obtained in $\chi^{RPA}_{22}(\vec{q},\omega)$. $\Omega$ decreases with the increase of $\vert V_{CDW}\vert$, and the normal state remains stable until $\Omega =0$ 
where CDW state starts to emerge. $\Im\chi^{RPA}_{22}(\vec{q},\omega)$ as a function of $\vert V_{CDW}\vert$ is plotted in Fig. \ref{fig:chirpa22}.
We notice that the critical interaction strength for CDW can be obtained by $\vert V^c_{CDW}\vert = 1/\chi^0_{22}(0,0)$ which is roughly proportional to $1/\Lambda^2$. 
This dependence of $\Lambda$ comes from the diverging density of states for the interband scattering in $\chi^0_{22}(\vec{q},\omega)$, which has also been observed in previous study.\cite{wangz2013}
We emphasize that the gap-like feature in $\Im\chi^0_{22}(\vec{q},\omega)$ is a direct consequence of the chirality of the Weyl nodes.
As a result, the undamped delta-function-like peak in the $\Im\chi^{RPA}_{22}(\vec{q},\omega)$ in the normal state is a unique feature for the Weyl semimetal, and it is rarely seen in other 
materials with CDW instability in which the peak is significantly damped by the particle-hole continuum.

\begin{figure}
\includegraphics[width=2.9in]{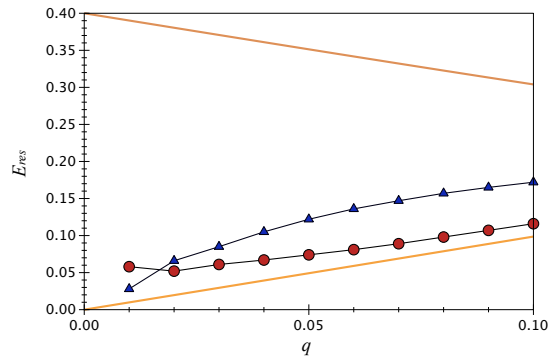}
\caption{\label{fig:plascdw} The frequencies of the two branches of collective excitations due to the mixture of the plasmon excitation and CDW phonon (axion) mode.
The parameters used are $\mu_{\pm} = 0.2$, $m_1 = 0.04$, $m_2=0$, $4\pi e^2/\kappa = 1$, and $\Lambda=5\mu$.
The solid lines represent the particle-hole continuum edges for the intraband ($\omega_{PHE_{intra}}$)and interband ($\omega_{PHE_{inter}}$) scatterings respectively.
The branch marked by the circle has more plasmon characters  while the one marked by the triangle has more CDW phonon (axion) characters.
Both branches should show up in the density-density correlation functions around $\vec{q}\approx 0$ and $\vec{q}\approx \vec{Q}_{CDW}$, which in principle can be detected by M-EELS.}
\end{figure}

{\it Resonant plasmon-axion excitation in CDW state} --
In the CDW state, we find two new effects in the bare susceptibilities. First, since the CDW order parameter mixes up the states with different chiralities, all the scatterings have finite 
probability. As a result, $\Im\chi^0_{11}(\vec{q},\omega)$ develops new spectral weights within the interband particle-hole continuum ($\omega > \omega_{PHE_{inter}}$). 
These extra contributions in 
$\Im\chi^0_{11(22)}(\vec{q},\omega)$ push the plasmon frequency toward the lower edge of the particle-hole continuum, 
resulting in the plasmon frequency $\omega_{pl}(\vec{q})$ being more dispersive in the CDW state than in the normal state\cite{lvm2013,hofmann2015}. 
Second, because $\chi^0_{12}(\vec{q},\omega)$ is non-zero in the CDW state, 
the frequencies of the collective excitations are now determined by 
\be
{\rm Det}(\hat{I} + \hat{\chi}^0(\vec{q},\omega)\hat{V}_{\vec{q}}) = 0,
\label{plas}
\ee
which has {\it two} solutions as explained below. Eq. \ref{plas} can be rewritten as:
\be
K_{11}(\vec{q},\omega)K_{22}(\vec{q},\omega) - v_{\vec{q}}V_{CDW}\vert \chi^0_{12}(\vec{q},\omega)\vert^2 = 0,
\ee
where
\beq
\label{k1122}
K_{11}(\vec{q},\omega)&=& 1+v_{\vec{q}} \chi^0_{11}(\vec{q},\omega)\approx A(1-\frac{\omega^2_{pl}(\vec{q})}{\omega^2}),\\
K_{22}(\vec{q},\omega)&=& 1+V_{CDW} \chi^0_{22}(\vec{q},\omega)\approx B(1-\frac{\omega^2_{pho}(\vec{q})}{\omega^2}),\nn
\eeq
and $\omega_{pl}(\vec{q})$ ($\omega_{pho}(\vec{q})$) is the plasmon (CDW phonon) frequency without $\chi^0_{12}(\vec{q},\omega)$.
Note that we have expanded $K_{11}$ and $K_{22}$ in the region without uncorrelated particle-hole excitations, i.e., $\omega_{PHE_{inter}} < \omega <\omega_{PHE_{intra}}$.
This is also the region where the undamped collective excitations could exist.
Substituting Eq. \ref{k1122} into Eq. \ref{plas}, it can be proved straightforwardly that two collective excitations exist with frequencies near $\omega_p$ and $\omega_{pho}$ respectively.
Since these two collective excitations are mixtures of the plasmon excitation and the CDW phonon mode, they should result in prominent peaks in both $\Im\chi^{GRPA}_{11}(\vec{q},\omega)$
and $\Im\chi^{GRPA}_{22}(\vec{q},\omega)$, emerging as resonant modes between these two channels.
Fig. \ref{fig:plascdw} plots the frequencies of these two resonant modes solved from Eq. \ref{plas} for different $\vec{q}$ together with $\omega_{PHE_{intra}}$ and $\omega_{PHE_{inter}}$.
Note that because magnitude of the spectral weight due to the interband scattering generally scales with $\Lambda^p$ ($p>1$), we could not draw a quantitative conclusion on the amount of 
the spectral weight for each resonant mode.
Nevertheless, the qualitative features of the collective excitations are robust regardless of the $\Lambda$ chosen.

It is remarkable to see that different mechanism to break the chiral symmetry in a Weyl semimetal could result in different behaviors of the plasmon excitation. 
Zhou {\it et al.} studied the plasmon excitation 
of a Weyl semimetal with the chiral symmetry broken by the application of a pair of the electric $\vec{E}$ and the magnetic $\vec{B}$ fields.\cite{hosur2015,zhouj2015}
In that case, the chirality-dependent chemical potential $\mu_\chi$ is introduced, which leads to a $\vec{q}$-dependent damping effect of the plasmon excitation. 
However, since $\mu_\chi$ does not provide any coupling between different Weyl nodes, the plasmon excitation does not have the resonant nature as discussed in the CDW state. 

It can be seen that the CDW phonon mode has a dispersion of $\omega(\vec{q}\to 0)\to 0$ as predicted by the Goldstone theory. However, the reason why we obtain the gapless Goldstone mode is that 
the effective Hamiltonian in Eq. \ref{h0} has the continuous translational symmetry. If the lattice effect is included, we will obtain a gapped CDW phonon mode.
Nevertheless, the Weyl semimetals discovered recently\cite{xus2015,wengh2015} have quite small $\vert\vec{Q}\vert$, 
thus the wavelength of the CDW state $\lambda_{CDW}\sim 1/\vert\vec{Q}_{CDW}\vert$ is much longer than 
the lattice constant. The lattice effect can therefore be argued to be quite small, and the CDW phonon mode should have a very small gap.
Moreover, because the phase of the CDW order parameter is shown to be the axion field\cite{wangz2013}, the CDW phonon mode is in fact a collective excitation of the axion field.
As a result, measuring these resonant plasmon-axion excitations around $\vec{q}\approx 0$ and $\vec{q}\approx \vec{Q}_{CDW}$ could serve as a reliable way to detect the axion field.

{\it Conclusion} --
In this paper, we have investigated the charge excitations of a Weyl semimetal in the charge-density-wave (CDW)state. We have found that in the normal state, 
an undamped collective mode could emerge in the density-density correlation function at the CDW ordering wavevector $\vec{Q}_{CDW}$ if there is an attractive interaction favoring the formation 
of the CDW state. In the CDW state, we have found that the plasmon excitation becomes more dispersive compared to the behavior in the normal state. Moreover, 
we have shown that due to the Umklapp scattering enabled by the CDW state, the plasmon excitation couples to the CDW phonon (axion) mode, leading to two branches of collective excitations. 
Both collective excitations can be seen in the density fluctuations at $\vec{q}\approx 0$ and 
$\vec{q}\approx \vec{Q}_{CDW}$ simultaneously, exhibiting an interesting resonant nature between two distinct channels. 
We have proposed that measuring the resonant plasmon-axion excitations around $\vec{q}\approx 0$ and $\vec{q}\approx \vec{Q}_{CDW}$ by M-EELS could serve as a reliable way to detect the CDW state as well as 
the nature of the axion field in a Weyl semimetal.
 
{\it Acknowledgement} -- This work is supported by a start up fund from Binghamton University.

\newpage
\section{Supplementary Materials}
\subsection{Matrix Elements for bare susceptibility matrix}
Following Eq. 6 in the main text,
the time-ordered bare susceptibility is a $2\times 2$ matrix $\hat{\chi}^0_{i,j}(\vec{q},\omega)$ defined as,
\be
\hat{\chi}^0_{i,j}(\vec{q},\tau)=\langle \rho_i(\vec{q},\tau)\rho^\dagger_j(\vec{q},0)\rangle,
\ee
where the density operators are defined in Eq. 3 in the main text.
First, we perform the Wicks decomposition and then convert the $c$ operators to the band basis,
\be
c_{R\sigma}(\vec{k},\tau)=\sum_{\alpha}U_{R_\sigma\alpha}(\vec{k})a_{\alpha}(\vec{k},\tau).
\ee
After we perform a Fourier transformation to Matsubara space, sum over the internal Matsubara frequency, and do the analytic continuation, we obtain the final expression of
retarded bare susceptibility as
\begin{widetext}
\beq
 \hat{\chi}^0_{i,j}(\vec{q},\omega)&=&-\sum_{\alpha,\beta}\sum_{\vec{k}}M_{i,j}^{\alpha,\beta}(\vec{k},\vec{q})\big(\frac{n_F(E_\alpha(\vec{k}+\vec{q}))-n_F(E_\beta(\vec{k}))}{\hbar\omega + i\delta
+E_\alpha(\vec{k}+\vec{q}) - E_\beta(k)}\big),\nn\\
M_{11}^{\alpha,\beta}(\vec{k},\vec{q})&=&\left(|\sum_{\sigma}U^{*}_{R_\sigma\alpha}(\vec{k}+\vec{q})U_{R_\sigma\beta}(\vec{k})|^2+|\sum_{\sigma}U^{*}_{L_\sigma\alpha}(\vec{k}+\vec{q})U_{L_\sigma\beta}(\vec{k})|^2\right),\nn\\
M_{22}^{\alpha,\beta}(\vec{k},\vec{q})&=&\left(|\sum_{\sigma}U^{*}_{L_\sigma\alpha}(\vec{k}+\vec{q})U_{R_\sigma\beta}(\vec{k})|^2\right)\nn\\
M_{12}^{\alpha,\beta}(\vec{k},\vec{q})&=&\sum_{\sigma}\left(U^{*}_{L_\sigma\alpha}(\vec{k}+\vec{q})U_{L_\sigma\beta}(\vec{k})+U^{*}_{R_\sigma\alpha}(\vec{k}+\vec{q})U_{R_\sigma\beta}(\vec{k})\right)\sum_{\sigma'}\left(U_{R_{\sigma'}\alpha}(\vec{k}+\vec{q})U^{*}_{L_{\sigma'}\beta}(\vec{k})\right)
\eeq
\end{widetext}
Since the density matrix is Hermitian we have $\hat{\chi}^0_{21}(\vec{q},\omega)=\big[\hat{\chi}^0_{12}(\vec{q},\omega)\big]^*$.

%\bibliography{weyl}{}

\end{document}